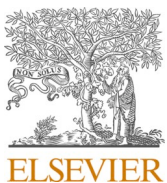
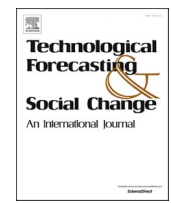

# Managing a blockchain-based platform ecosystem for industry-wide adoption: The case of TradeLens

Marin Jovanovic [a],[*], Nikola Kostić [b], Ina M. Sebastian [c], Tomaz Sedej [b],[c]

[a] *Department of Operations Management, Copenhagen Business School, Solbjerg Plads 3, 2000 Frederiksberg, Denmark*
[b] *Department of Accounting, Copenhagen Business School, Solbjerg Plads 3, 2000 Frederiksberg, Denmark*
[c] *Center for Information Systems Research (CISR), The MIT Sloan School of Management, 245 First Street, Cambridge, USA*

A B S T R A C T

The proliferation of blockchain-based platform ecosystems in recent years has prompted scholars across various disciplines to explore the conditions leading to their successful deployment. However, developing a blockchain-based platform ecosystem creates various challenges for the platform sponsor that may influence industry-wide adoption and, ultimately, the platform's success. This study follows the development of TradeLens, a leading global shipping platform ecosystem underpinned by blockchain technology. We examine the factors affecting industry-wide adoption among global supply chain actors by unpacking platform value drivers and platform governance mechanisms identified at TradeLens. While the platform value hinges on the digitalization of workflows and the ecosystem leverage, the platform governance includes strategic (off-chain), technology (on-chain), and interoperability (on- and off-chain) governance – as mechanisms for effectively managing a blockchain-based platform ecosystem. This paper contributes to the literature on blockchain-based platform ecosystems and the platform literature.

## 1. Introduction

"*I won't mince words here – we do need to get the other carriers on the platform. Without that network, we don't have a product. That is the reality of the situation.*"
- Marvin Erdly, Head of TradeLens.

In recent years, blockchain and distributed ledger technologies have evolved from their origins in digital currencies (e.g., Bitcoin, Ethereum) to various other use cases, including blockchain-based platforms for digital transformation with the prospect of disrupting established supply chains (e.g., pharma, food, logistics), governments (e.g., El Salvador) and possibly society in its entirety (e.g., disintermediation of trust) (Chen and Bellavitis, 2020; Halaburda et al., 2022; Lacity and Van Hoek, 2021). One of the most exciting blockchain applications has appeared in global supply chains (cf. Agi and Jha, 2022; Kouhizadeh et al., 2021; Kumar et al., 2020; Pournader et al., 2020; Wamba and Queiroz, 2020). While the global container shipping supply chain accounts for approximately 60 % of all seaborne trade, valued at 12 trillion USD (Statista, 2021), the shipping industry is plagued by inefficiencies (Balci, 2021; Hastig and Sodhi, 2020; Yang, 2019). These inefficiencies stem predominantly from manual and time-consuming customs clearance and contracting processes, outdated legacy systems with data trapped in silos, and disparate perspectives on the transaction state that typically involve more than 40 actors (Jensen et al., 2019, 2018).

To address these issues, incumbent shipping companies and startups have begun to leverage blockchain technology to facilitate the emergence of platform ecosystems that would create value across the global shipping supply chains (Chin et al., 2022; Kumar et al., 2020; Pereira et al., 2019). Specifically, blockchain-based platform ecosystems offer decentralized solutions for digitizing the records of container transiting history that can handle data security with great precision (Jensen et al., 2019; Schmeiss et al., 2019). The unique feature of a blockchain-based platform ecosystem is that it can provide services to global supply chain actors through seamless connections without having access to their sensitive data (Cai et al., 2021; Chang et al., 2020). Hence, digitizing the physical workflows with a blockchain-based platform ecosystem can reduce transaction costs by making transactions more transparent and seamless (Babich and Hilary, 2020; Gaur and Gaiha, 2020). Furthermore, a blockchain-based platform ecosystem generates a data layer (Alaimo and Kallinikos, 2021), that can be further utilized through a marketplace of complementary applications that create additional value for all platform contributors, complementors, and users (e.g., smart-contract solutions for automating and self-executing transactions) (Bonina et al., 2021; Dolgui et al., 2020; Schmidt and Wagner, 2019). However, developing a blockchain-based platform ecosystem creates unique challenges for the platform sponsor which may influence the industry-wide adoption necessary for the platform's success (Helfat and Raubitschek, 2018; Uzunca et al., 2022).

* Corresponding author.
*E-mail addresses:* mjo.om@cbs.dk (M. Jovanovic), nk.acc@cbs.dk (N. Kostić), isebasti@mit.edu (I.M. Sebastian), tse.acc@cbs.dk (T. Sedej), tsedej@mit.edu (T. Sedej).

First, while the general benefits of blockchain-based platforms for the global supply chain actors have been explored in the literature (cf. Garg et al., 2021; Pereira et al., 2019; Rejeb et al., 2021; Wamba and Queiroz, 2020), the value-creating opportunities of a blockchain-based platform ecosystem for different public and private global supply chain actors remain underexplored (Jensen et al., 2019). Moreover, the platform value is often unevenly distributed among global supply chain actors, difficult to define, and hard to measure (Bauer et al., 2022; Goldsby and Hanisch, 2022). For instance, blockchain features may bring to global supply chain actors improved inter-organizational workflow management (Kostić and Sedej, 2022), increased operational efficiency (Toorajipour et al., 2022), and opportunities to leverage the open ecosystem (Cenamor and Frishammar, 2021; Eisenmann et al., 2009; Schmeiss et al., 2019). Therefore, it is essential for platform sponsors who are actively building an industry-wide blockchain-based platform ecosystem to comprehensively understand the platform value from the perspective of diverse global supply chain actors (Kouhizadeh et al., 2021).

Second, effective platform governance mechanisms are critical to ensure industry-wide adoption and platform success (Murthy and Madhok, 2021; Rietveld and Schilling, 2021; Uzunca et al., 2022). A blockchain platform architecture, by design, puts in place on-chain technology governance features (Lumineau et al., 2021; Wareham et al., 2014). However, the underlying blockchain features are often insufficient to establish the level of trust that would convince prospective global supply chain actors to join the platform (De Filippi et al., 2020; Zhao et al., 2022). Consequently, a platform sponsor needs to design various off-chain governance mechanisms that would address the inter-organizational concerns of prospective actors (Kostić and Sedej, 2022; Schmeiss et al., 2019; Uzunca et al., 2022). Still, issues concerned with establishing platform interoperability and the associated process of digital standardization (Rossi et al., 2019), and, more broadly, platform regulation (Cusumano et al., 2021; Jacobides and Lianos, 2021) are underexplored in the platform literature. These concerns are all connected to lowering the barriers to platform ecosystem adoption and harnessing the potential of blockchain technology (Kumar et al., 2020). Overall, the interplay of different governance mechanisms at nascent stages of blockchain-based platform ecosystem development is underexplored (cf. Chen et al., 2022).

This study seeks to understand ***how a platform sponsor manages a blockchain-based platform ecosystem in the shipping industry to ensure industry-wide adoption***. The study follows the development of TradeLens, a blockchain-based platform ecosystem jointly developed by Mærsk Line,[1] a logistics conglomerate, and IBM, a multinational information technology company. TradeLens is a leading blockchain-based platform ecosystem in the shipping industry that aims to establish an ecosystem that is connected end to end, embracing various global supply chain actors, such as shippers, cargo owners, ports, and freight forwarders. To this end, TradeLens has established a shared and secured blockchain-based platform ecosystem that allows global supply chain actors to significantly diminish the costs of information exchange and create an open marketplace of applications. Analysis of the rich qualitative data unpacked the platform value drivers, namely the digitalization of workflows and the ecosystem leverage. Additionally, the study delineated three platform governance mechanisms, namely strategic (off-chain), technology (on-chain), and interoperability (on- and off-chain) governance. These value drivers and governance mechanisms collectively facilitated the effective management of TradeLens. The paper contributes to the literature on blockchain-based platform ecosystems and, more broadly, the platform literature.

The paper is structured as follows. First, we develop our theoretical orientation. Second, we introduce our research methods before presenting the findings of our study. Third, we discuss the results and position our contribution within the broader platform literature. Finally, we highlight the managerial implications and limitations of our study.

## 2. Theoretical background

### 2.1. Blockchain-based platform ecosystems

The shipping industry has traditionally relied on the physical movement of large amounts of paper documentation, which has been associated with delays, human error, and fraud (Stopford, 2009). In addition, the digital infrastructure (Fürstenau et al., 2019) in the shipping industry is disseminated across different legacy systems developed by ports, freight forwarders, terminal operators, and customs offices that use different formats and standards (e.g., EDI, fax, e-mail) to exchange documents about the physical shipment (Toorajipour et al., 2022). More than 40 organizations can share data and documents in a single shipment, producing large amounts of documentation in diverse formats (Jensen et al., 2018; UN, 2016). More importantly, the manual handover of paper documents is largely inefficient, expensive, and prone to error, which increases the price of the overall shipment and produces supply chain bottlenecks (Sund et al., 2020). Indeed, the processing of accompanying trade documentation can be even more expensive than moving an actual shipment (Chang et al., 2020).

To address these issues, both practitioners and scholars propose developing a blockchain-based platform ecosystem to securely connect the public and private global supply chain actors (Jensen et al., 2019). Such platform ecosystems are particularly important for low-trust industries, such as shipping (Stopford, 2009). The underlying blockchain features, combined with the platform ecosystem features, are increasing the platform value (Cennamo, 2021) and the likelihood of industry-wide adoption (Goldsby and Hanisch, 2022).

First, a blockchain-based platform ecosystem uses the platform architecture with a distributed ledger system, which maintains the records (or transactions) in a chain of blocks (Schmeiss et al., 2019). A distributed architecture guarantees security, transparency, and control among transaction partners, allowing ecosystem participants to transfer digital assets or business-relevant information across firm boundaries through a shared, tamper-proof distributed ledger (Kumar et al., 2020; Pereira et al., 2019). Therefore, blockchain features address the trust issue in the industry by moving some of the complexity-related issues from the organizational to the technical level (Beck et al., 2018; Catalini and Gans, 2020). In other words, instead of entrusting a platform sponsor with potentially sensitive data, blockchain features offer a technical solution for handling the distribution and integrity of shared data (Jensen et al., 2019). Moreover, scholars suggest that blockchain platform architecture can reduce costs associated with supply chain transactions by decreasing information and search costs as well as expenditure on post-contractual control (Kostić and Sedej, 2022). Therefore, an immutable ledger permits automated actions and performance tracking among contractual partners (Schmidt and Wagner, 2019). Finally, Lumineau et al. (2021) suggest that data integrity and reliability, enabled by the blockchain, can better detect opportunism while reducing monitoring costs.

Second, apart from leveraging the blockchain features, platform sponsors are expanding the platform value by developing complementary applications and engaging with external developers, thereby creating an open platform ecosystem (Cenamor and Frishammar, 2021;

---

[1] To ease the exposition, in the remainder of the paper Mærsk Line will be referred to simply as Mærsk. When referring to Mærsk Line's parent company we use the term Mærsk Group.

Pereira et al., 2019). Indeed, secure and transparent data exchange among the business-to-business platform ecosystem actors holds the potential to generate an open marketplace of complementary applications (Jovanovic et al., 2021). More importantly, in industries where interdependence among actors is high (e.g., shipping industry, phone network industry, computer operating systems), the presence of complementors could significantly drive the platform value (e.g., movies on a streaming service, applications for a smartphone, shipping industry) (Ozalp et al., 2018). Therefore, the network externalities of a platform could represent a large portion of its overall platform value (Cennamo, 2021; Jovanovic et al., 2021; Ozalp et al., 2018). However, when a large portion of a platform value is derived from its network externalities (i.e., the size of the end-user network and the size of the complementor network) (Cennamo, 2021), new technology, such as blockchain, may not be able to displace an incumbent technology, even when the benefits brought forth by the new technology compared to the old technology are well understood (Lacity and Van Hoek, 2021). Both complementors and end users make platform adoption decisions based on which platform they believe will have the largest number of ecosystem actors. Consequently, ensuring that platform adoption is industry-wide is of utmost importance for a platform sponsor (Eisenmann, 2008; Eisenmann et al., 2009). Therefore, characterizing the platform value drivers of a successful blockchain-based platform ecosystem may provide clues to the factors critical for the platform's success (Yoffie et al., 2020; Zhu and Iansiti, 2019).

*2.2. Governing a blockchain-based platform ecosystem*

While it is widely understood that the success of a business-to-business platform ecosystem depends on industry-wide adoption (Jovanovic et al., 2021), less is known about how a platform sponsor employs different governance mechanisms to attract prospective contributors, complementors, and end users in a blockchain-based platform ecosystem (Beck et al., 2018; Bonina et al., 2021). First, a platform sponsor decides on a set of blockchain features (e.g., private permissioned blockchain) that define the level of security, transparency, and control over the shared data (Lumineau et al., 2021). Second, the initial capital investment may grant the platform sponsor a wide market presence, the ability to attract ecosystem actors, and the potential to dominate the market even if the technology was initially considered inferior (Kim and Mauborgne, 2019; Ozalp et al., 2018). Third, the literature argues that the neutral position of a platform sponsor has to be established ex ante in order to signal that the platform ecosystem goals are aligned with the industry-wide interests of ecosystem actors (Jensen et al., 2019). Fourth, managing an industry-wide platform ecosystem requires addressing cooperation and competition among the actors (Cennamo and Santaló, 2019; Hannah and Eisenhardt, 2018). Therefore, the literature argues that the initial specification of governance terms is critical for the success of the platform ecosystem (Goldsby and Hanisch, 2022; Uzunca et al., 2022). In particular, defining the governance terms early on is acutely important when deploying a blockchain-based platform ecosystem because certain collaboration and coordination rules are embedded in the initial specification and are enforced according to the business logic encoded in the blockchain software (Chen et al., 2021). Fifth, a platform sponsor needs to engage with the prospective ecosystem actors to assist them in transitioning to a new way of exchanging and validating information and integrating their existing legacy systems into the blockchain-based platform ecosystem (cf. Monika and Bhatia, 2020; Pólvora et al., 2020). Overall, how the blockchain features and platform ecosystem governance interact (Lumineau et al., 2021; Zhao et al., 2022) and how the platform sponsor engages with and onboards the potential ecosystem actors are concepts that remain insufficiently researched in the platform literature (Jovanovic et al., 2021).

Finally, a platform sponsor needs to decide on the interoperability governance in relation to complementary and competing platforms (Marsal-Llacuna, 2018). For instance, ecosystem actors are reluctant to risk investment in a specific blockchain-based platform ecosystem if the selected platform does not look likely to "take off" (Jensen et al., 2019; Markus et al., 2006). While (digital) standardization is a proven solution to safeguard customers, the roadmap is lengthy (DCSA, 2020). Therefore, it is typical for platform sponsors to actively participate in the ongoing process of developing a digital standard (DCSA, 2022). However, platform interoperability governance – specifically, for blockchain-based platform ecosystems – is less explored in the literature (cf. Wimmer et al., 2018).

## 3. Research method

*3.1. Research design*

We conducted the study using an exploratory in-depth case of TradeLens, a world-leading example of a large-scale, blockchain-based platform ecosystem that provided a rich source of information for this study. We opted for a case study research design (Yin, 2017) because it could provide valuable insights of a particular phenomenon in depth and within its real-life setting. Moreover, this research design was particularly useful since the primary objective of this paper was to identify the most relevant factors for the effective management of TradeLens – namely, platform value and platform governance mechanisms – that would influence the decision of various global supply chain actors to join TradeLens. Given the dearth of studies offering empirical evidence on blockchain-based platform ecosystems, this study follows a qualitative theory-building approach (Eisenhardt and Graebner, 2007). In particular, qualitative data are a source of well-grounded, rich descriptions and are able to explain processes in identifiable local contexts (Miles and Huberman, 1994).

Moreover, we followed an abductive approach (Dubois and Gadde, 2002). Our empirical fieldwork, case analysis, and theoretical framework development occurred simultaneously. In such studies, the initial theoretical framework is continuously adjusted, as a result of both unexpected empirical findings and theoretical insights attained during the process. An abductive approach can result in useful cross-fertilization in which new combinations are created through the amalgamation of well-established theoretical models and new concepts derived from the collection of empirical data (Dubois and Gadde, 2002).

*3.2. Data collection*

Between February 2018 and September 2021, the data were collected from three sources: i) in-depth semi-structured interviews; ii) informal conversations with individuals involved with TradeLens; iii) secondary data including TradeLens's documentation, industry reports, industry conference presentations, news articles, press releases, and participation in industry conferences and live webinars. A semi-structured interview guide was developed that sought to unpack the platform value drivers and governance mechanisms for the effective management of TradeLens that would trigger industry-wide adoption (Fontana and Frey, 1998). When conducting in-depth semi-structured interviews, the authors had a list of theory-based questions and themes to cover. However, they did not follow a rigid order of questions and allowed interviewees to describe the phenomenon in their own terms and from their own viewpoints (Kvale, 1996). To obtain a preliminary understanding and overview of TradeLens, initial exploratory interviews with Mærsk and GTD Solution/TradeLens in 2018. Although TradeLens went live in December 2018, and despite all the potential upside, the adoption of the platform remained sluggish. Consequently, in 2019, additional interviews were conducted with Mærsk and a sample of other industry participants to understand the reasons for the slow uptake. During these interviews, particular attention was paid to the factors holding back the industry-wide adoption of the blockchain-based platform ecosystem and to exploring how the two founding companies intended to address them. Additionally, an interview with a prominent shipping industry analyst was conducted to obtain an external perspective on the issues in the container shipping sector in general, as well as specific concerns related to blockchain-based platforms in the industry. In 2020 and 2021, additional interviews were conducted with global supply chain actors, including terminals, ocean carriers, and customers. In total, the authors conducted 24 interviews with key informants that were recorded and transcribed using the automated speech-to-text software service. In addition, detailed notes were taken during and immediately after each interview. Table 1 provides an overview of the conducted interviews.

The informants were identified using natural or convenience sampling (Collis and Hussey, 2014) because the choice of participants was influenced by interviewees' roles and their involvement in TradeLens. The data were collected from informants in several large companies, including the two largest ocean carriers in the world (Mærsk and MSC, respectively), a technology provider (IBM), container terminals (GCT, ICTSI terminals, and APM terminals) and large customers/exporters (e. g., AB InBev and Van den Ban Autobanden B.V.). This approach was deemed necessary because blockchain-based platform ecosystems such as TradeLens typically require complex interactions between multiple organizations. The majority of informants held senior positions in their respective companies (e.g., CEO, CIO, CTO, VP, head of department). They were chosen because the nature of their positions allowed them to provide a high-level view of the decision-making process at TradeLens. Another informant selection criterion filtered only TradeLens ecosystem partners who were already involved in the process of joining TradeLens. Repeated interviews allowed us to crosscheck information collected from other respondents and secondary data.

In addition to formal interviews, several informal talks were held with individuals involved with TradeLens. These include the CEO of GTD Solutions/TradeLens, Head of Digital Business Solutions at Port of Rotterdam (a TradeLens member), and the CIO of Hapag Lloyd AG (a TradeLens member). In addition to primary data, secondary data were collected to complement and verify interview data. Secondary data used in this study consists of TradeLens's published documentation, industry reports, industry conference presentations, news articles, and press releases. Discrepancies between interview data and secondary data raised new questions, which guided subsequent data collection and analysis (Alvesson, 2011). Appendix A maps the industry conferences and live webinars, and Appendix B provides an overview of secondary data sources.

### 3.3. Data analysis

Interview transcripts were coded using constant comparative analysis (Corbin and Strauss, 1990; Glaser and Strauss, 2017). As the research developed and new data were collected, the categories identified were continuously compared to the existing data (Alvesson, 2011). When data produced novel or contradictory information, the categories were adjusted to take these new developments into account. This process was repeated until no new categories emerged and no new information was inconsistent with existing categories (i.e., until theoretical saturation was reached) (Corbin and Strauss, 1990). This constant comparative analysis involved data triangulation by cross-checking statements across informants and verifying them against secondary data. Initial open coding produced eleven empirical themes describing platform value drivers and twelve empirical themes indicating different underpinnings of effective platform governance mechanisms that influence the decision of global supply chain actors to join TradeLens. As we cycled between data collection, coding, and existing theory, initial empirical themes were grouped into five conceptual categories using axial coding (Corbin and Strauss, 1990; Strauss and Corbin, 2015). These were, in turn, synthesized into two aggregate dimensions – namely, the blockchain-based platform ecosystem value composed of improving the digitalization of workflows and ecosystem leverage, and the blockchain-based platform ecosystem governance composed of strategic ecosystem governance, technology ecosystem governance, and interoperability governance. The data structure that resulted from this iterative analysis is presented in Fig. 1. The following section discusses each of the dimensions in greater detail.

**Table 1**
The overview of conducted interviews.

| # | Date | Type | Position | Company | Location |
|---|------|------|----------|---------|----------|
| 1 | 14.2.2018 | Interview | Controller and Analyst | Gefion Insurance | Case site (Gefion) |
| 2 | 7.3.2018 | Interview | Controller | Gefion Insurance | Case site (Gefion) |
| 3 | 2.5.2018 | Interview | Digital Product Manager | Mærsk | Case site (Mærsk) |
| 4 | 24.5.2018 | Interview | Lead IT Architect | GTD/TradeLens | Case site (GTD/TradeLens) |
| 5 | 14.6.2018 | Interview | Special Consultant & Chief Consultant | Ministry of Industry, Business and Financial Affairs | Ministry of Industry, Business and Financial Affairs |
| 6 | 3.7.2018 | Interview | Digital Product Manager | Mærsk | Case site (Mærsk) |
| 7 | 6.7.2018 | Interview | Head of Data and Business Development | Danish Maritime Authority (DMA) | DMA |
| 8 | 7.3.2019 | Interview | Senior Consultant | IBM | Online/Zoom |
| 9 | 14.3.2019 | Interview | Digital Product Manager | Mærsk | Case site (Mærsk) |
| 10 | 4.7.2019 | Interview | Global Head of Integration | APM Terminals | Case site (Mærsk) |
| 11 | 10.10.2019 | Interview | CEO, Partner | SeaIntelligence Consulting | SeaIntelligence Consulting |
| 12 | 21.10.2019 | Interview | Digital Product Manager | Mærsk | Case site (Mærsk) |
| 13 | 30.3.2020 | Interview | Digital Product Manager | Mærsk | Online/Zoom |
| 14 | 31.3.2020 | Interview | Head of Strategy and Operations | GTD/TradeLens | Online/Zoom |
| 15 | 20.5.2020 | Interview | CDIO (MSC); Chairman (DCSA) | MSC/DCSA | Online/Zoom |
| 16 | 26.5.2020 | Interview | Project (Stream) Lead at Global International team | Anheuser-Busch InBev | Online/Zoom |
| 17 | 26.5.2020 | Interview | Vice President, Blockchain Solutions | IBM | Online/Zoom |
| 18 | 10.6.2020 | Interview | President/CEO | Global Container Terminals Inc. | Online/Zoom |
| 19 | 7.7.2020 | Interview | Various departments | Pacific International Lines | E-mail |
| 20 | 3.9.2020 | Interview | CTO | Youredi | Online/Zoom |
| 21 | 9.9.2020 | Interview | CIO | YILPORT holding | Online/Zoom |
| 22 | 27.5.2021 | Interview | CIO | International Container Terminal Services, Inc. | Online/Zoom |
| 23 | 1.7.2021 | Interview | Customs & Trade Compliance Manager | Van den Ban Autobanden B.V. | Online/Zoom |
| 24 | 14.9.2021 | Interview | Program Director | DCSA | Online/Zoom |

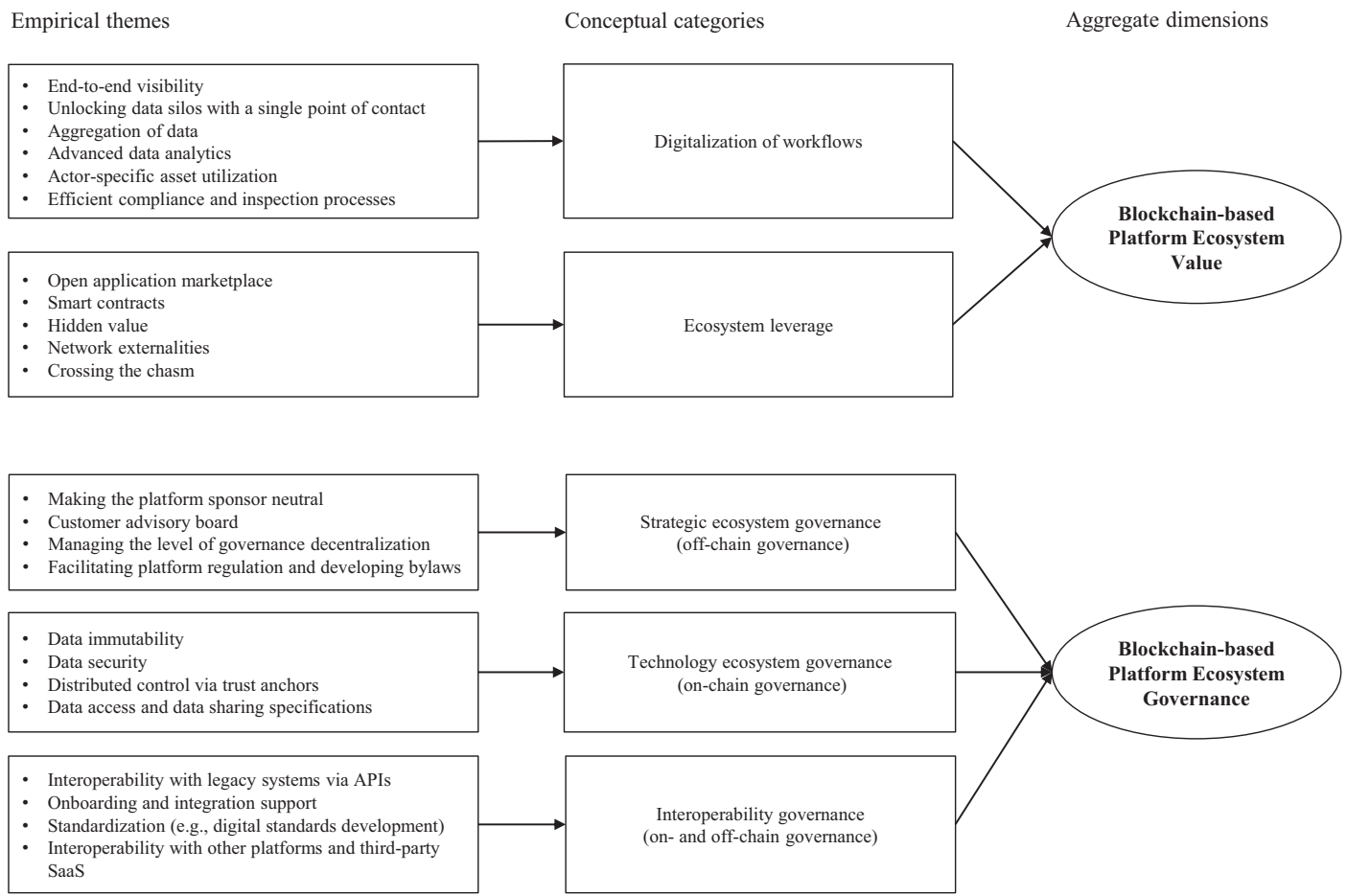

Fig. 1. Data analysis structure.

## 4. Findings

### 4.1. TradeLens – a blockchain-based platform ecosystem

In order to drive platform adoption, TradeLens leadership needed to demonstrate the platform's value. TradeLens provided an efficient, standardized, and secure exchange of information and trade documentation that could bring *digitalization of workflows* to the shipping industry and unlock substantial financial gains by addressing massive inefficiencies. Additionally, digitizing and automating the exchange of trade documentation would allow companies to redirect their efforts from handling burdensome administration to more value-adding activities. Moreover, TradeLens has created the foundation for *ecosystem leverage* with a planned marketplace of innovative complementary services that can benefit actors in addition to a high level of network externalities and interdependency among the global supply chain actors in the shipping industry.

#### 4.1.1. Digitalization of workflows in global supply chains

TradeLens creates value from providing *end-to-end visibility,* which involves the entire shipping ecosystem. For many shipping firms, the key benefit was that TradeLens would enable them to easily gather information on the status of the cargo at each stage of the journey. The ability to have visibility into the previous blind spots in their journey was the dominant factor that led to their decision to opt for TradeLens. This sentiment was echoed by the project leader of the Global International team at Anheuser-Busch InBev who argued:

"*The problem [with] platforms such as INTTRA, GTN or CargoSmart, for example, is [that they are] communication platforms between us as a shipper and our freight provider or carrier. There's no other supply chain party posting data in these platforms. So, these platforms have a lot of data, but the data are only coming from the carriers or shippers. That's where most of the platforms are still failing today – they don't have the full ecosystem, all of the supply chain or cargo transport participants under one platform*."

TradeLens promises significant improvements in operational efficiencies for global supply chain actors. This study finds that the back offices of organizations in the shipping industry spend significant amounts of time on gathering, exchanging, and reconciling inter-organizational information due to the data that is trapped in organizational silos. Since data is shared in several different formats, standardized and digitized trade documentation could *unlock data silos* and remove these inefficiencies by automating cross-organizational information exchange. This, in turn, will result in significant reductions in back-office workload and associated costs, further allowing organizations to re-focus on more value-adding activities. A senior Mærsk executive provided an interesting anecdote:

"*Today, a lot of the data is tied up in organizational silos and those silos don't talk to each other. I remember, I had a chat with someone in the network who said, 'We are all connected, only not to each other.' I thought in the beginning it was funny, but actually it is the case. We think we are connected, and we somehow are, but often what happens in shipping, today, is that people are connected by fax, email, EDI, all types of communication, that are really not up to speed with what is required today, which is real time information. The value proposition in the beginning [of the TradeLens project] was simply to take out the friction, remove the friction from the communication that goes on between the* *different parties in the supply chain, because it is a chain, so no actor can act independently anyway.*"

TradeLens aims to establish itself as *a single point of contact* in the digitalized shipping industry. The benefits of maintaining a single interface with many different global supply chain actors in the shipping industry were noted by several respondents. The president and CEO of Global Container Terminals Inc., for instance, observed:

"*[The benefit of TradeLens is that] we only need to connect to the platform, and all the shippers need to connect to the platform versus us connecting to each of the 15 shippers directly, which costs a lot of money, takes a lot of time, and does not create a lot of value.*"

Through digitizing trade documentation, structuring data pipelines, enabling real-time information, and automating multi-party interactions, TradeLens laid the foundations for unlocking digital value from data-driven innovations. The chief digital and information officer (CDIO) at Mediterranean Shipping Company (MSC) observed that the *aggregation of data* can improve existing services:

"*I think by digitizing or digitalizing, we can bring a little bit more value, maybe monetize different things that we're not doing today and overall being more efficient. We could even aggregate some data from the platform and improve it.*"

Interviewees from Mærsk and GTD Solution further contend that accurate, reliable, and near real-time information is available for TradeLens ecosystem members. They can improve their internal planning systems by employing near real-time *advanced data analytics*, which can result in reduced uncertainty about the availability of goods and a consequent strengthening of their capacity to serve their clients. In addition to improved customer service, advanced data analytics can enable ocean carriers to minimize costs related to moving empty containers. As observed by the vice president of Blockchain Solutions at IBM:

"*One of the biggest problems carriers face is empty container position. They move a lot of empty containers from one part of the world to the other because goods need to be shipped from one part of the world. To solve that optimization problem, the industry loses tens of billions of dollars. They spend billions of dollars in moving stuff unnecessarily. They want to minimize that. So, data that's flowing through TradeLens will be valuable for that too because it just tells you the total location of all the containers that are available globally.*"

"*Data about the actual movement of the goods, instead of that coming in delayed. Because you know, from an ocean carrier, what they care about is when the vessel leaves the port and when it enters into the next port. So, all the information that they give out is based on their part of the journey. The same goes for the terminal. The terminal cares about when the container enters the terminal and when it leaves again. For the trucker, it's about when they pick up the container and when they drop it off at the gate. So, you know, instead of some human stitching up that whole patchwork of information, the idea is that the infrastructure will do that automatically.*"

Next, TradeLens has facilitated an increase in *actor-specific asset utilization*. However, the benefits of asset utilization may differ among ecosystem actors. For instance, governments may benefit from timely and accurate information in relation to customs inspection and audits, ports and terminals may improve on-site throughput, freight forwarders may maximize the utilization of their trucks and synchronize their operations according to dynamic changes in the global supply chain, and ocean carriers may optimize their vessel utilization. A senior IBM consultant provided an example of how TradeLens digital value may come in different forms. He noted:

"*The primary benefit is to increase the utilization of the assets that they [platform ecosystem participants] own. So, for a terminal, it's about speeding the throughput of vessels, so that you can make more money from your investment in the terminal, the reason that we can speed that throughput is because we are increasing the efficiency of the data exchange, and we're making it more reliable.*"

"*Different members of this platform get very different benefits. That's also quite a challenge when it comes to communicating the value, that these sort of platforms bring value that is different for governments to the value for an importer, an exporter, and different to the value for a terminal.*"

Moreover, partnerships with customs authorities are major source of advantage and value for prospective members of TradeLens because faster and more *efficient compliance and inspection processes* are an important part of improving operational efficiencies for the entire shipping industry. Several interviewees see these partnerships as both essential and mutually beneficial. By joining TradeLens, customs authorities are able to obtain all their required documentation in an efficient, standardized, and reliable manner. Since TradeLens is underpinned by blockchain, authorities can benefit from the tamper-proof log of transactions that creates a reliable audit trail. Additionally, customs authorities can benefit from a single interface (i.e., with TradeLens) for all members that have already onboarded the platform ecosystem. A senior IBM consultant explained the benefits of TradeLens for the authorities:

"*One of the benefits is that they [the authorities] can improve their targeting selection and risk assessment processes. So, they can't possibly inspect every single movement of goods across the world, they have to decide what to inspect. You do that based on data. And there are currently two challenges with that data. They don't necessarily get data as early as they want, and they go through many steps. So, before it [the data] gets to the government, lots of unintentional errors are made and lots of information is lost in that process.*"

"*Governments are interested in any and all data that they can get in order to facilitate trade and sort out the bad guys and claim their revenue. So, we plan to be an enabler of data for regulators*."

In a similar way, partnering with customs authorities has enabled TradeLens to offer additional value for prospective ecosystem members from easier and faster *compliance processes*. Digitizing trade documentation on TradeLens can lead to an easier compliance process because TradeLens ensures that documents are standardized and compliant with the current rules and regulations of particular jurisdictions. This may be especially valuable for smaller players in the ecosystem who do not possess sufficient resources to adjust their local documentation to the changing demands from regulators. An example was given by a project leader of the Global International team at Anheuser-Busch InBev:

"*The stamp [on documents for some customs authorities] has to be in black ink. If it's in blue, they don't accept it. All of this takes a lot of effort for a shipper to remember, and we sometimes forget to put the ink in black and they seize the container. The container is stuck in the port, where we pay $150 storage a day. We need to send them a new bill of lading, but via post. So, you can already do the math how much this little mistake will cost us. If we have a functioning and trusted platform in place, the amendment of the bill of lading is immediate. It's within minutes, not days, so issues like this will not exist anymore. That's where we will save the money.*"

### 4.1.2. Ecosystem leverage

Another way in which TradeLens seeks to encourage innovation is by developing an *open application marketplace* (e.g., application store) with open APIs that allow ecosystem members to build fit-for-purpose services on top of the platform. There are currently three applications available in the application marketplace, all developed by TradeLens. The first is TradeLens Core, a supply chain management tool, which captures shipment information and delivers it to relevant ecosystem participants via API or a user interface. The second application is the electronic Bill of Lading (eBL), which is an electronic version of a legal and commercial document between a shipper and a carrier that specifies the type, quantity, and destination of the goods. The third application is a Bill of Lading Verifier, which is tailored specifically for banks and other financial institutions, enabling more effective trade finance processes. The leadership of TradeLens believes that these applications will become the main source of the company's revenue in the long run. While TradeLens developed all three of the initial applications, the aim of an open application marketplace is to invite complementors to develop their own innovative applications and continually contribute to the usability and versatility of the platform ecosystem. The vice president of Blockchain Solutions at IBM described these considerations:

> "*Third parties will come in simply because, 'Hey, you guys have gathered an ecosystem of network members and clients and you're just pumping data through the system, how can I add value to let's say a shipper, by giving them a better invoice dispute resolution solution or an application?' The raw data can be used in various ways, and that is where third parties come in to deliver more value.*"

> "*We have called it an app store; marketplace is another word you can use. We are in conversation with others who are also having ideas for apps, and they can vary from logistics, but they can also go into the financial sector. We're not going to have all of the ideas and certainly there are some things that we can't do, like financing, for example. But we can facilitate those activities through TradeLens.*"

Interviewees also expressed the belief that incentivizing third parties to develop new, innovative solutions will contribute to an expansion of the TradeLens ecosystem by offering unique solutions previously unavailable to global supply chain actors, such as ports and terminals. As noted by the president and CEO of Global Container Terminals Inc.:

> "*If the TradeLens Marketplace will become an economic clearing hub for rail volume consolidation or truck volume consolidation and, through that, they will reduce price to our customers, the shippers, we would want to be a part of that.*"

Further, the blockchain-based architecture of TradeLens allows for the development of *smart contracts* – programmable algorithms stored on the blockchain that are executed when predetermined conditions are met. A senior consultant from IBM described an example of how they see the value of TradeLens's smart contracts:

> "*[…] The capability of smart contracts is to agree that certain processes are enforced between parties, or a certain set of rules are enforced before the transaction is accepted on the blockchain. So, for example, the ability to coordinate the approval processes of certain regulatory documents, like sanitary certificates, or these kind of things, where you've got a government authority that issues it and signs it, and then somebody else has to approve it. And then in another country, another government authority looks at it and approves it. We're working on streamlining these processes by using the smart contract functionality.*"

Finally, some of the potential value of TradeLens is relatively *"hidden"* in the early stages, meaning that TradeLens will need to convince potential adopters that the platform ecosystem will deliver more value in the future. A Mærsk executive explained its ambition and resolve to leverage TradeLens and unlock the value they cannot envision yet. He explained:

> "*We think that probably there will be a lot of things, which we cannot think of right now, which will pop up. So, we really want to add builders to come with creative ideas. Disruptive power, implicitly, because if you make things transparent, the middlemen have to change their role. They have to find ways to upgrade the service and then the whole work starts to shift.*"

Attracting prospective global supply chain actors is key to creating value in a blockchain-based platform ecosystem. Since actors in the shipping industry are highly interdependent, *network externalities* of TradeLens are high and TradeLens can only reach its full potential if all relevant parties participate in the same platform ecosystem. Since dozens of large organizations, such as ocean carriers, ports, terminals, and governmental authorities, are involved in a single container shipment, any missing nodes would require TradeLens customers to engage individually with the prospective ecosystem member in order to complete the flow of information. Additionally, the participation of all ecosystem actors involved in a shipment journey increases transparency and helps alleviate the invoicing concerns that characterize contemporary global supply chains. The vice president of Blockchain Solutions at IBM described these problems:

> "*If you move X number of containers from one part of the world to the other, you've got to pay a bill. The bill is basically made up of multiple parts. You've got to pay the inland trucking lanes on either end, you've got to pay the ocean carrier, you've got to pay the port for storage of the container. So, the bill is pretty hard to understand. Moreover, you have no idea whether you're paying for the right services. With TradeLens, you could see that [the container] was at this port for this many days. It was on this bunker journey at this point in time, bunker oil surcharges were at this point in time. So, just tracking that helps you deliver a better invoice.*"

> "*Let's say port terminal operator says, I don't want the ports to know what's in the container, but the port says, well, I have to put it on the train in my operations, I have to do something with it, and I also want to know. We talk through those supply chain groups, and they very quickly realize there's so much benefits that they step over that.*"

TradeLens demonstrated its ability to "*cross the chasm*" more quickly than its competitors. Since Mærsk, the biggest ocean carrier in the world at the time, was the initiator of TradeLens, it was able to leverage its dominant market position to onboard downstream and upstream members of the value chain. In particular, TradeLens onboarded five of the top six global ocean carriers. Our data suggest that, while competing ocean carriers were initially reluctant to join TradeLens, there was a considerable appetite from other actors in the shipping industry to adopt the platform. Interviewees from Mærsk and IBM noted that this initial interest contributed to the perceived value of the TradeLens network in two material ways. First, as the platform sponsor and a heavyweight actor in the shipping industry, Mærsk was able to leverage its own networks to onboard new members. Second, the growing ecosystem created an incentive for other ocean carriers to join the network, as several of their customers and partners were already participants. They had the incentive to do so because the benefits they could realize would expand as the number of ecosystem actors that joined TradeLens increased. A senior consultant from IBM explained the logic behind creating a critical mass of shipping transactions on the platform. He explained:

> "*The reason we launched this platform with Mærsk is that by launching a trade platform with a large carrier, we're able to overcome some of the challenges around the network effects for the platform. You need a critical mass of people and information on a platform in order for it to deliver value. By starting and seeding the platform with the information from Mærsk, we already have nearly 20% of global trade or containers that are traveling around the world available through that platform. So, the thinking was to bring together a technology company [IBM] and a big ocean carrier [Mærsk]. That way, you can create the critical mass that you need to make this successful.*"

This reasoning was confirmed by a digital product manager at Mærsk when asked about plans for TradeLens. He observed:

> "*The idea was to find the strong partners that could create critical mass, and then have them influence their networks to grow the [TradeLens] network.*"

## 4.2. Managing off-chain and on-chain governance on a blockchain-based platform ecosystem

Due to the high competitiveness and low trust in the shipping industry, governance decisions were critical in influencing the willingness of global supply chain actors to adopt TradeLens. Because the TradeLens platform ecosystem is shared across multiple actors including rivals, effective governance decisions helped align goals and incentives, ensured intellectual property protection, and mitigated disputes. Based on the findings, TradeLens's governance is discussed on three distinct levels. The first level involves decisions on ownership, participation, distribution of benefits, and platform development. In other words, it encompasses agreements and arrangements between global supply chain actors pertaining to the overall goals and objectives of the platform ecosystem. For the purposes of this paper, this level of governance is termed *strategic ecosystem governance* (or off-chain governance). The second level refers to the *technology ecosystem governance* (or on-chain governance). It involves the technical execution of blockchain platform architecture and sets blockchain agreements between global supply chain actors. This level defines who is in control of the data and ensures that shared data are protected. Finally, the third level involves *interoperability governance* (on- and off-chain governance), including managing the interoperability with legacy systems of ecosystem actors and with other shipping platforms (blockchain and non-blockchain). The notion of interoperability governance is mainly related to the concepts of efficiency and risk.

### 4.2.1. Strategic ecosystem governance (off-chain governance)

With regard to strategic ecosystem governance, TradeLens had to signal to prospective ecosystem members that the platform ecosystem was open and neutral, and that it would establish and continuously manage an advisory board to represent interests of all ecosystem members. The first step for Mærsk was creating a separate business entity within Mærsk Group, called GTD Solution, that was to operate at arm's length from the rest of Mærsk and treat Mærsk as any other ecosystem member. GTD Solution ensures that the terms offered to a particular ecosystem member, regardless of who it is, are equivalent to the terms available to other ecosystem members, such as large ocean carriers. The aim of establishing GTD Solution was to signal that a *neutral platform sponsor* operated TradeLens, and that its goal was to unlock digital value across the entire industry, rather than accruing benefits to Mærsk alone. Respondents from Mærsk anticipated that neutrality would lead to an increase in the perceived trustworthiness of the platform, resulting in industry-wide adoption. The head of strategy and operations at GTD Solution/TradeLens believed that this approach would alleviate concerns of prospective ecosystem members – in particular, competing ocean carriers. He noted:

> "When you interact with GTD, you're interacting with it as if it were its own company, even though it's part of the Mærsk group. And so, all of a sudden you no longer have the risk of data that you're giving as a competitor to Mærsk, to this platform, getting in the hands of Mærsk itself because there is a separation of system, separation of people, separation of legal constructs."

The second step was establishing the TradeLens *customer advisory board*, making sure that decisions on the development of the platform, such as data standardization and the product roadmap, are transparent and aligned with other members of the ecosystem. The idea behind creating an advisory board was described by a digital product manager at Mærsk:

> "We have created the customer advisory board whose job it is to represent those who are TradeLens customers, so that we make decisions through collaboration. So, for example, around data standardization, the product roadmap, and things like that to make sure that what we're doing is aligned with the mission statements that we have for the company. But at the end of the day, every decision that we make is done where we think will be aligned with what is good for the fullest supply chain ecosystem."

> "The advisory board was a response to the many questions about: 'Why should I join something that IBM has built for Mærsk?' The idea with the advisory board was simply to listen to the industry, to set up a team of people who would be representative of different actors in the supply chain, and to open that up. I think that, on the one side, it was a good move to have a voice of more than just one customer. At the same time, I think it illustrates one of the problems in shipping in general that, on the surface, we all trust each other but, actually, we don't. That is what the advisory board was set down to do, to be the voice of the whole industry and to also have an influence on the development."

Although Mærsk and IBM made initial decisions on the functioning of the platform ecosystem, in order to drive adoption, the two companies needed to take a step back and involve other global supply chain actors. This approach signified a move toward a more decentralized governance model, which aims to ensure that decisions are transparent and aligned with the members of the entire industry, rather than residing with a few dominant players. Executives from Mærsk and IBM suggested that the final shape of the advisory board is not yet determined. As the platform ecosystem grows, TradeLens needs to *manage the level of governance decentralization*. This may require tradeoffs between the inclusion of a larger number of ecosystem participants and flexibility in terms of decision making and TradeLens development. This tradeoff was also noted by the CEO and partner at SeaIntelligence Consulting, a leading shipping industry expert:

> "If you invite a lot of carriers and give them veto right, you run into the INTTRA[2] problem – when you reach a minimum threshold, some of the carriers will be happy, and it's impossible to move it further. The other extreme would be to position it as a Mærsk–IBM project, get flexibility, but immediately alienate all the other carriers."

While there seems to be a consensus across respondents that the inclusion of customs authorities will create considerable value for the ecosystem as a whole, some interviewees noted that getting more of them to join might present an administrative challenge for TradeLens. Several customs authorities around the globe still require original paper documentation, including stamps and signatures. Fully realizing the benefits of TradeLens will, therefore, require some *changes in platform regulations and developing bylaws*. Moreover, adopting an industry-wide platform is largely a political decision because many governments are skeptical of sharing data through a global platform. A program director at DCSA explained:

> "So, bylaws could help interoperability, but it's not only for interoperability. Let's imagine for a second a world in which interoperability never happens in terms of eBL solution providers; to us, it still makes sense [to create standard bylaws]."

### 4.2.2. Technology ecosystem governance (on-chain)

Concerning technology ecosystem governance, TradeLens had to address the issues of data immutability, data security, distribution of control via trust anchors, and data access and sharing models on the blockchain-based platform ecosystem. First, the inherent attributes of blockchain technology ensure the *data immutability* feature of TradeLens. An informant from IBM elaborated on the importance of data immutability in platform ecosystem governance, stating:

> "Because of the consensus algorithms, we can say that, once some data is in there, it's impossible for a single organization to change it. And that has

---

[2] INTTRA is an EDI-based portal for standardized contained bookings that was founded as a joint venture of seven of the world's largest ocean carriers in late 2000.

*some very useful consequences when you are looking at scenarios where you've got a lack of trust. So, either between the private sector and government, or between different private sector participants where serious amounts of money are at stake. So, I think that the ability, through an appropriate consensus algorithm, to say that no party can unilaterally change the history is powerful."*

A similar point was highlighted by the project leader at AB-InBev who provided an example:

*"It's a trustful source because we see the digital signature, and we see it's not a changeable field. But if you use normal kind of EDIs or IDocs or whatever, you can post the data to someone's system, and he can of course amend it, and it is happening. It is happening also in our work. It's time to change it. It's the 21$^{st}$ century. We don't want to work with paper. We don't want to create hard copies of documents and have them signed, stamped, no."*

Similarly, the underlying blockchain platform architecture is an important element contributing to TradeLens *data security*. However, the level of security and control over data depends on decisions made during the design of the blockchain platform architecture – notably the number and ownership of validating nodes. Respondents from Mærsk and IBM noted that, even though the two companies initiated TradeLens, it quickly became apparent that companies could not run all, or even the majority, of the blockchain nodes because this would result in excessive power in the hands of a single company, which would run against their positioning as a genuinely neutral industry-wide platform ecosystem. The head of strategy and operations at GTD Solution/TradeLens described these efforts:

*"We took a number of steps to address concerns like, 'What if I give my data to this platform that is controlled by my key competitor? What can they do with those insights, and what might they do with it?' Or, 'How can I be sure that whatever I do here, another shipping line isn't going to get an advantage from mine being on there that I can't realize?'"*

A blockchain is a tool for *distributing control* over a shared ledger among multiple participants in which no single participant can unilaterally exercise full authority over the system. TradeLens offered participating ocean carriers the option to host and manage a blockchain node. Carriers who opt for this option are referred to as "*trust anchors*" and maintain an exact copy of the ledger. Trust anchors are known to the network by their cryptographic identities. They participate in the consensus mechanism, meaning that they validate transactions, host data, and assume a critical role in securing the network. As opposed to public permissionless blockchains, in which the ledger is replicated on every node in the system (i.e., universal data diffusion), only trust anchors hold a copy of the ledger on TradeLens's private permissioned blockchain. A Digital product manager at Mærsk explained:

*"The trust anchors are handling all the verification on behalf of the number of participants in the network. That, of course, speeds up everything because you don't replicate the same truth on so many nodes…plus then…less transactions, faster speed. The idea is that instead of having everyone setting up a node, you save the cost of doing that. Both in terms of the transaction and also money, of course. Because you don't need that server running, you don't need expertise in running that server and setting it up, and making sure everything is in place, security, privacy, GDPR and so on."*

Finally, in terms of *data access*, TradeLens leveraged a permissioned blockchain structure to develop the "*data sharing specification*". It is based on a data model, referred to as a "consignment hierarchy", which TradeLens adopted from the United Nations Centre for Trade Facilitation and Electronic Business (UN/CEFACT). It differentiates between three layers – shipment, consignment, and equipment. A shipment defines a commercial and financial relationship between a buyer and a seller and typically includes the buyer's and the seller's banks. A consignment is the operational execution of that commercial relationship and includes more organizations, such as ports and inland transportation providers, but will tend not to involve banks. Equipment is the unit that contains goods that are part of the commercial relationship (e.g., a shipping container). There is a many-to-many relationship between these three layers. A shipment, for example, can be comprised of several different consignments, and a single consignment can have many shipments associated with it. TradeLens identifies a role that a particular ecosystem actor can play in executing a shipment, consignment, or equipment. The identified role, in turn, determines what transactions a particular actor can access.

More importantly, in addition to the data sharing specification, TradeLens also leverages IBM blockchain *"channel architecture"*, which specifies how data is shared within the blockchain environment. A channel is established for each participating ocean carrier, and information is distributed only to those nodes participating in a channel (i.e., multi-channel diffusion). This means that none of the customer information from ocean carriers will be distributed to other ocean carriers that are not a part of a specific transaction. Moreover, only the hash value of commercially sensitive information is stored on the blockchain, so that only authorized participants are able to see if the information has changed (through a changed hash value), without actually seeing the underlying data. Documents are stored on a single node, and are accessed at runtime by other nodes on a particular channel as permissions allow. This means that participants in the network stay in control of their own data, while TradeLens handles the operational integration of these independent ecosystem actors using standard protocols. TradeLens's *data retention policy* defines how long the data is stored on the platform. The hashes on the blockchain ledger are stored indefinitely, but the hash cannot be used to recreate the original data. An executive from TradeLens explained:

*"It's important to remember that the data resides with the data owner. The platform as such doesn't hold the data. The platform enables the different actors to exchange data when it is relevant so if you are a party relevant to a shipment, you have access, if you have been delegated that access by the data owner. If I'm a producer and I'm shipping sneakers from Bangladesh to the US, I know my suppliers, I know my forwarder, I know my ocean carrier, and I delegate access to all of them to be able to help me out on creating the trail of information that is needed for me to get it somewhere."*

*"We have a data sharing model, called a diagram or table, where we arranged who can see what. Basically the objective is that the sharing is minimized amongst the parties who handle a specific container that they see the data… so that the transparency is where the benefits mostly come from. If I have the container, and I send it to you, if we all know what's in it, and when it's coming, and when it should be coming, the rest of the world doesn't have to know. And if you are involved in 200 containers, you'll see of those 200 containers, that information. So, that limits the visibility to a very effective level."*

#### 4.2.3. Interoperability governance (on- and off-chain governance)

In interoperability governance, TradeLens had to address interoperability with legacy systems, providing onboarding and software integration support, digital standardization, and interoperability between TradeLens and other platforms. First, as noted by respondents from Mærsk and GTD Solution, there are two different options for connecting to TradeLens. Prospective ecosystem members can interact with the platform by linking it to their proprietary IT systems via open application programming interfaces (APIs). *Integrating legacy systems with TradeLens* via *APIs* is particularly attractive for major organizations who are processing large amounts of data. As the head of strategy and operations at TradeLens observed:

*"[…] if you think of a shipping line or a terminal, the amount of data that they are using is considerable, and those data need to be in our platform at*

*scale. And that's not going to happen if you have to have 40 people in a room typing data into a system that can then go to TradeLens."*

*"Of course, many of these different organizations also have to change their tech set up to accommodate to working with the APIs from TradeLens and to make sure that this goes into their operational systems because one thing is that they're feeding data into the system. They also need to somehow consume the APIs and be able to use that operationally so it is not just saying, 'We all know how this is working, we can just open up the gates and everything will just start flowing.'"*

*"So, API is the next level from an EDI connection. And the reason why that's important is that everybody knows that EDI connections are very hard to change. It takes a lot of work, and it is very expensive, whereas an API is way more flexible."*

An ecosystem member, Anheuser-Busch InBev, also reiterated the importance of interoperability with the legacy systems as noted by its project leader:

*"Imagine that you're shipping over 250,000 containers a year. We need to have a platform directly interfacing with our system. Because the operational teams cannot go and manually put all the data needed for ocean booking or shipping instructions. That would require an army of people, and we definitely don't want this. We must have something connected."*

Ecosystem members can also take the option of employing a user interface instead of integrating their existing systems with the platform ecosystem. This is particularly relevant for smaller firms who either do not have a core legacy system or consider such integration too expensive. Therefore, the level of integration and the amount of platform-specific investment are dependent on a specific actor, its size, and the amount of data exchanged, which requires the actor-specific approach to interoperability governance.

Next, TradeLens offers *onboarding and integration support* for prospective members to assist in integrating their legacy systems with TradeLens. The onboarding team provides guidance and assistance with data and process mapping and integration testing to ensure the current legacy systems of ecosystem members are able to both feed data and consume data from TradeLens. One TradeLens executive explained:

*"Full integrations will happen and then the data starts flowing because as of today it's actually only Mærsk's data that is available to all parties whereas the work of onboarding the rest of the carriers takes a lot of time because there is some data mapping going on."*

Open APIs and the support offered by GTD Solution and IBM reflect the ambition of the two companies to motivate actors to join the platform ecosystem by making connecting to TradeLens as easy as possible. Some respondents, however, noted that introducing new technology is a challenging endeavor that may not resolve persistent data quality issues. As observed by the CTO of Youredi, TradeLens's integration solutions provider:

*"When updating an existing integration from old batch-based to a modern API […] many players will give you the impression that using new technology will somehow magically fix problems with the old technology, and this is very often not the case. Implementing a new API might give a company the opportunity to fix issues with old technology, but new technology is no guarantee that issues will be fixed. Fixing data quality errors usually requires improvements to several downstream systems in the process."*

This can be particularly difficult for companies using a lot of customized software accumulated over the years. A digital product manager at Mærsk argued that:

*"Over time you build a lot of technical debt, and just by introducing new modern APIs, that's never going to solve the legacy problems, or the technical depth that you may have built up over 20 years."*

Next, an important topic related to interoperability is *standardization*, which remains a problem in the shipping industry. The aim is to achieve higher levels of standardization, adopt some of the most basic vocabulary conventions developed by UN/CEFACT, and leverage UN/CEFACT's Supply Chain Reference Data Model (SCRDM). Adopting this model helps to make the TradeLens platform interoperable with the systems of other ecosystem actors who have also adopted SCRDM. In addition to adopting the UN/CEFACT standard data model in 2018, TradeLens is also involved in co-creating industry digital standards by collaborating with the recently established Digital Container Shipping Association (DCSA). DCSA is a not-for-profit, independent organization founded by nine of the ten largest global ocean carriers in 2018 with the mission to establish standards for the entire shipping industry, which will allow systems to "speak to each other". Therefore, it will play a crucial role in establishing industry-wide interoperability. As the CDIO at Mediterranean Shipping Company (MSC) and the chairman of DCSA observed:

*"When they tell you a plane has landed, you understand that the plane has landed and it's going to the terminal. It's going to discharge the passengers. When you say a vessel has arrived, it means nothing. It could have arrived in berth, it could have arrived and be waiting outside, it could have arrived in the mouth of a river, and it still has 12 hours to get to the berth."*

Finally, *interoperability between TradeLens and other platforms* is a critical consideration for potential participants because it decreases switching costs and risks. In 2020, developers from Oracle, IBM, and SAP disclosed that they had completed cross-platform testing and were able to connect consortia of firms clustered on different platforms. Since TradeLens is run on an IBM blockchain and GSBN (a competing blockchain-based platform) is run on an Oracle blockchain,[3] this could well mean that the risk for partners to join either platform will be diminished considerably. The CDIO at MSC, for example, suggested:

*"The world needs more than one platform, that's for sure. There has to be interoperability because you will always have some of the parties using platform A and some of the parties using platform D or E or F, because that's going to develop."*

TradeLens has established data arrangements that enable firms to exchange data with a platform that is partially overlapping with TradeLens and, therefore, competes in one segment but is also a powerful data complementor in another segment. A senior IBM executive explained TradeLens's position in relation to the competing platforms:

*"If they're big enough, or interesting enough, we connect, for example, some are regional logistic platforms. What do we do? We make a data deal. We say, you give this information, you get that information from us, but we both win. And we have another level of digitization, some might be for our app store. So, for example, groups of platforms who are in the financial, around the bill of lading, there's a lot of scrutiny and checks and banks, guarantees and that type of stuff. Some have even a blockchain solution for that. So, we will plug that into TradeLens or work together. We think always that we set the entrance of the universe."*

Additionally, TradeLens ensured *interoperability with third-party Software-as-a-Service (SaaS)* providers who offered onboarding support to prospective clients. This is a particularly interesting option for companies who are hesitant to let onboarding teams from GTD Solution or IBM get too close to their proprietary data. As a result, the TradeLens leadership was considering the possibility of integrating TradeLens into SAP because this could create an additional incentive for ecosystem members to adopt the platform. As a Digital product manager at Mærsk noted:

---

[3] Both IBM blockchain and Oracle blockchain are powered by Hyperledger Fabric, an enterprise-grade permissioned distributed ledger framework, providing modular architecture.

*"[if we integrate with SAP] then you would have another gateway because all the companies using SAP would suddenly see: 'Oh, this is a component within my suite of ERP that I'm already working with.'"*

## 5. Discussion and conclusion

This study aims to unpack the most important factors in managing a blockchain-based platform ecosystem in the shipping industry. Our study extends, in several ways, the literature on platform ecosystems supported by blockchain technology in the context of global supply chains and the platform literature more broadly.

First, the study contributes to the emerging platform literature by exploring one of the largest active blockchain-based platform ecosystems operating globally (Goldsby and Hanisch, 2022; Jensen et al., 2019). It also contributes to the literature on platform adoption and the antecedents of platform success (Zhu and Iansiti, 2019) by unpacking the critical elements influencing the decision of global supply chain actors to embark on digital transformation by joining a blockchain-based platform ecosystem (Hsuan et al., 2021; Schmeiss et al., 2019).

Second, while blockchain-based platform ecosystems serve as a promising vehicle for digital transformation in data-sensitive industries (Lacity and Van Hoek, 2021), the potential value of a blockchain-based platform ecosystem that will legitimize industry-wide adoption needs to be clearly conceptualized and communicated (Garg et al., 2021; Jovanovic et al., 2021; Rejeb et al., 2021). The study shows that platform value is affected by both blockchain-based and platform ecosystem features. In particular, the digitalization of workflows has unlocked organizational and cross-organizational data silos (e.g., data islands) and allowed for efficient end-to-end collaboration across the entire global supply chain (Phadnis and Fine, 2017; Schmidt and Wagner, 2019). However, the value generated and prioritized may differ depending on the actor's type and role in the ecosystem (e.g., authorities, ports, ocean carriers) (Cennamo, 2021; Parida and Jovanovic, 2021). Additionally, the data layer allowed for the creation of a marketplace that can be opened to specialized complementors to develop innovative solutions that expand the platform ecosystem value (Jovanovic et al., 2021). High task-oriented interdependence among the global supply chain actors creates strong network effects (Gawer and Cusumano, 2014; Helfat and Raubitschek, 2018; Ozalp et al., 2018), making crossing the chasm to industry-wide adoption critical for the long-term success of a platform ecosystem (Clarysse et al., 2014; Moore, 2014).

Third, the study unpacks a set of key governance mechanisms for a blockchain-based platform ecosystem – strategic, technology, and interoperability governance. More specifically, due to the complex interplay between the blockchain-related governance and inter-organizational governance features, our study makes a distinction between on-chain and off-chain governance (cf. digital and non-digital governance features, Chen et al., 2022). Strategic ecosystem governance safeguards platform ecosystem members from potential opportunistic behavior by other members or the platform sponsor. Studies have shown that platform sponsors use signaling to establish the role of a trusted intermediary or a trusted platform sponsor (Deng et al., 2021). This is particularly important for the industry-wide adoption of a blockchain-based platform ecosystem that connects "frenemy" actors (Adner et al., 2019). In this study, we show that establishing a neutral position of a platform sponsor positively affects industry-wide adoption, similar to standard-setting organizations and open-source communities (O'Mahony and Karp, 2022). In our study, establishing an advisory board assured potential adopters that decisions on operations and future development of the platform ecosystem will be governed by a broader ecosystem (Pombo-Juárez et al., 2017). The study also contributes to the literature on participation architecture (West and O'Mahony, 2008) by highlighting the challenge of setting the right level of advisory board decentralization (Lacity and Van Hoek, 2021).

Fourth, this study finds that the platform sponsor must proactively work on the platform regulation (Cusumano et al., 2021; Jacobides and Lianos, 2021) and create bylaws for paperless trade because current laws and regulations do not adequately address the legal framework for digital and, especially, blockchain-based trade (Goldby, 2016). This research contributes to the emerging literature on interoperability governance (Wimmer et al., 2018) by underscoring both interoperability with different legacy systems and competing platforms (Ziolkowski et al., 2020) as vehicles to increase the perceived usability of the platform but also to decrease the risk of platform lock-in and platform-specific investments (Ozalp et al., 2018). Similarly, the study highlights the importance of digital standards to strengthen blockchain-based platform adoption (Zachariadis et al., 2019).

Finally, the study provides a perspective on technology ecosystem governance (Lumineau et al., 2021; Wareham et al., 2014) in the context of a blockchain-based platform ecosystem. In particular, this research has demonstrated that design choices of the blockchain platform architecture have important implications for governance and platform adoption (Lumineau et al., 2021). More specifically, the study provides evidence on emerging models to balance data access and permission management (Azaria et al., 2016). Our research also provides a rare example of smart contract application in the context of the global supply chain (Agrawal et al., 2022; Murray et al., 2021; Ziolkowski et al., 2020).

### 5.1. Managerial implications

Despite limitations to generalizability based on a single case, insights from this study provide valuable information for firms in other industries, which are developing or exploring blockchain-based platform ecosystems. Platform sponsors should carefully consider proposed design decisions when building a blockchain-based platform because these decisions will determine the platform characteristics (e.g., security, transparency) and influence the adoption decisions of other actors in the industry. Demonstrating value and building an appropriate governance structure help align the interests of various actors and promote the diffusion of an industry-wide blockchain-based platform ecosystem, irrespective of the particular industry. In addition, establishing a collaborative structure such as TradeLens's customer advisory board can ensure the collective participation of representative members of heterogeneous groups, further increasing the likelihood of industry-wide platform adoption. Finally, ensuring the interoperability of a blockchain-based platform with legacy systems, as well as with other (potentially competing) blockchain-based platforms, can lead to increased efficiency and lower risk, which can incentivize a greater number of global supply chain actors to adopt a particular platform.

### 5.2. Limitations

This study is subject to several limitations. The case study method is potentially vulnerable to researcher-induced bias during both data collection and analysis. Even though case-based research and qualitative data facilitate the investigation of complex phenomena, they also restrict the statistical generalizability of findings (Yin, 2017). This study does not suggest that platform value and platform governance fully explain all aspects of effective management of a blockchain-based platform ecosystem and that achieving positive results in each of these areas will unequivocally result in industry-wide platform adoption. The latter will depend on several contingencies of each specific implementation project, such as individual preferences of a particular global supply chain actor, the prior relationship with a platform sponsor, and specifics of platform regulation. Yet, the two identified dimensions outline the most important considerations of current TradeLens members and potential adopters, influencing their decision to adopt the platform. TradeLens will need to find a balance between addressing the requirements of various global supply chain actors and the complexity to meet all these specific requirements. Finally, some regulatory hurdles are inhibiting TradeLens from reaching its full potential. To achieve the goal to simplify compliance processes, several customs authorities and governments worldwide would need to start accepting data from TradeLens.

### 5.3. Future research

Future studies should continue to explore the TradeLens case as it develops and evaluate the progress made in each of the two dimensions. Researchers could also explore the TradeLens case in depth in specific areas (e.g., the financial impact of joining TradeLens for a particular company, the effect of standardization on the success of the platform, impact of digitized trade documentation and automatic execution of pre-defined rules on management control). We propose that researchers apply the two identified dimensions – platform value and governance – to explore how firms build successful blockchain-based platform ecosystems in other industries. As the emerging landscape of blockchain-based platform ecosystems continues to evolve, we expect it will fundamentally change how companies collaborate, compete, share data, and develop innovative products and services.

### Data availability

No data was used for the research described in the article.

### Appendix A. The overview of conferences and webinars

| Date | Type | Title | Organizer | Location |
| --- | --- | --- | --- | --- |
| 04.11.2017. | Conference participation | Nordic Blockchain conference | ITU Copenhagen | Copenhagen |
| 18.04.2018. | Conference participation | Blockchain conference and exhibition | Blockchain Expo World Series | London |
| 18.06.2019. | Conference participation | TOC Europe | TOC Events Worldwide | Rotterdam |
| 11.11.2019. | Conference participation | SHIP TECH 2019: Conference on the future of shipping | ShippingWatch | Copenhagen |
| 19.02.2020. | Webinar | Learning about DCSA's Track & Trace standards | Digital Container Shipping Association (DCSA) | Online |
| 12.05.2020. | Webinar | Digitalisation and data standardization: time for the maritime industry to act | Maritime Optimization and Communication | Online |
| 26.05.2020. | Webinar | Adapting to 'New' New Normal: The Impact of COVID 19 | TOC Events Worldwide | Online |
| 09.06.2020. | Webinar | Accelerating Digitalization: The Role Of Start-up Tech In Post-covid-19 Supply Chains | TOC Events Worldwide | Online |
| 03.07.2020. | Webinar | Where next for global shipping? | Executive MBA in Shipping and Logistics: Copenhagen Business School | Online |
| 14.07.2020. | Webinar | Global Overview of the Container Shipping Market | Intermodal Digital Insights | Online |
| 15.07.2020. | Webinar | Global Smart Container Forum | Intermodal Digital Insights | Online |
| 05.08.2020. | Webinar | An electronic bill of lading, considered the holy grail of the maritime industry | IBM Blockchain/TradeLens | Online |
| 12.08.2020. | Webinar | How 3PLs and FFWs move from linear logistics to a platform business model | IBM Blockchain/TradeLens | Online |
| 19.08.2020. | Webinar | BiTA + TradeLens: Alignment & Opportunities Moving Forward | FreightWaves | Online |

### Appendix B. The overview of the secondary data sources

| Outlet | Webpage |
| --- | --- |
| TradeLens webpage | https://www.tradelens.com/ |
| TradeLens blog | https://www.tradelens.com/blog |
| TradeLens press releases | https://www.tradelens.com/blog/all-press-releases |
| TradeLens documentation | https://docs.tradelens.com/ |
| GTD Solution webpage | https://www.gtdsolution.com/ |
| Digital Container Shipping Association (DCSA) | https://dcsa.org/ |
| JOC.com (Container shipping and trade news and analysis) | https://www.joc.com/ |
| Coindesk | https://www.coindesk.com/ |
| Ledger Insights | https://www.ledgerinsights.com/ |
| Wired | https://www.wired.co.uk/ |
| World Economic forum | https://www.weforum.org/ |
| LinkedIn posts | https://www.linkedin.com/ |
| Twitter Posts | https://twitter.com/ |
| IBM Blockchain | https://www.ibm.com/blockchain |
| PWC | https://www.pwc.com/gx/en/industries/technology/blockchain/blockchain-in-business.html |
| Coin Telegraph | https://cointelegraph.com/ |
| The Loadstar | https://theloadstar.com/ |
| Container news | https://container-news.com/ |
| SeaIntelligence Consulting | https://www.seaintelligence-consulting.com/ |
| Supplychain dive | https://www.supplychaindive.com/ |
| Global Trade review | https://www.gtreview.com/ |
| Globe newswire | https://www.globenewswire.com/en |
| Logistics Middle East | https://www.logisticsmiddleeast.com/ |
| Seatrade Maritime News | https://www.seatrade-maritime.com/ |
| Port Technology | https://www.porttechnology.org/ |
| Express Computer | https://www.expresscomputer.in/ |

(*continued*)

| Outlet | Webpage |
|---|---|
| Container Management | https://container-mag.com/ |
| The Maritime Executive | https://www.maritime-executive.com/ |
| BTC Manager | https://btcmanager.com/ |
| PR Newswire | https://www.prnewswire.com/ |
| Splash247.com | https://splash247.com/ |
| Business Blockchain HQ | https://businessblockchainhq.com/ |
| Forbes | https://www.forbes.com/ |
| Market Research Reports | https://www.marketresearchreports.com/maritime |
| Harvard Business Review | https://hbr.org/ |
| MIT Technology Review | https://www.technologyreview.com/ |
| The National Law Review | https://www.natlawreview.com/ |
| Coin Rivet | https://coinrivet.com/ |

Dr. Marin Jovanovic is an assistant professor at the Department of Operations Management at Copenhagen Business School and a visiting scholar at Luleå University of Technology. He received a Ph.D. degree in industrial economics and management from the KTH Royal Institute of Technology and a Ph.D. degree (cum laude) in industrial management from the Universidad Politécnica de Madrid. His research has been published in academic journals, such as Organization Science, Journal of Product Innovation Management, R&D Management, International Journal of Operations & Production Management, Technovation, International Journal of Production Economics, Journal of Business Research, and Industrial Marketing Management. His research interests include the digital transformation of manufacturing and maritime industries, platform ecosystems in the business-to-business context, and artificial intelligence. Marin has held positions at the ESADE Business School and the University of Cambridge.

Dr. Nikola Kostić is a researcher at the Department of Accounting at Copenhagen Business School. His research has been published in a leading accounting journal Accounting Horizons. His research interests include accounting and management control issues in inter-organizational settings, digital standardization and digital transformation, inter-organizational implications of blockchain technology use, and digital infrastructure governance.

Dr. Ina M. Sebastian is a Research Scientist at the MIT Center for Information Systems Research (CISR). Her research areas are partnering and value creation in digital ecosystems, and digital transformation of large enterprises. Prior to joining MIT CISR in 2014, Ina completed a Ph.D. with a focus on Information Systems Management at the Shidler College of Business, University of Hawaii. In her dissertation research, she studied how Electronic Health Record Systems enable and constrain relational coordination in multidisciplinary clinician teams. One of her ongoing research interests is the role of digital capabilities for relational and adaptive coordination in health care systems, within and across organizational boundaries. Ina also holds an MBA from University of Hawaii and a Diplom in Business Administration from RWTH Aachen University.

Dr. Tomaz Sedej is an external lecturer at the Department of Accounting at Copenhagen Business School and a visiting research scholar at MIT Center for Information Systems Research. He holds a Ph.D. from Copenhagen Business School, MBA from Bradford University, School of Management and a B.Sc. from University of Ljubljana, Faculty of Economics. Before joining academia, he worked in a private sector, where he held several different positions at L'Oréal, Coca-Cola HBC and Gorenje Group Nordic. His research interests include blockchain, platform ecosystems, digital partnering and digital transformation.